\documentclass[intlimits,twoside,a4paper]{article}

\usepackage{amsmath,amssymb}
\usepackage{graphicx}
\usepackage{wrapfig}
\usepackage{textcomp}

\usepackage[T2A]{fontenc}
\usepackage[cp1251]{inputenc}

\usepackage[eqsecnum]{cmpj2}

\issue{2015}{18}{3}{33005}
\doinumber{10.5488/CMP.18.33005}

\title[Modelling the intermixing effects in highly strained asymmetric
InGaAs/GaAs quantum well]%
{Modelling the intermixing effects in highly strained asymmetric
InGaAs/GaAs quantum well}
\author[M. Souaf \textsl{et al.}]%
{M. Souaf\refaddr{label1}, M. Baira\refaddr{label1},
H. Maaref\refaddr{label1}, B. Ilahi\refaddr{label2,label1}\thanks{bilahi@ksu.edu.sa}}
\addresses{
\addr{label1} University of Monastir, Laboratory of Micro-optoelectronic and Nanostructures, Departement of Physics, Faculty
of sciences,  Tunisia
\addr{label2} King Saud University, Department of Physics and Astronomy, College of Sciences,  11451
Riyadh, Saudi Arabia
}

\authorcopyright{M. Souaf,  M. Baira, H. Maaref, B. Ilahi, 2015}
\date{Received April 18, 2015, in final form June 10, 2015}

\begin{document}

\maketitle

\begin{abstract}

In this work, we have theoretically investigated the intermixing
effect in highly strained In$_{0.3}$Ga$_{0.7}$As/GaAs quantum well (QW) taking into consideration the composition profile change
resulting from in-situ indium surface segregation. To study the
impact of the segregation effects on the postgrowth intermixing, one
dimensional steady state Schrodinger equation and Fick's second law
of diffusion have been numerically solved by using the finite
difference methods. The impact of the In/Ga interdiffusion on the QW
emission energy is considered for different In segregation
coefficients. Our results show that the intermixed QW emission energy
is strongly dependent on the segregation effects. The interdiffusion
enhanced energy shift is found to be considerably reduced for higher
segregation coefficients. This work adds a considerable insight into
the understanding and modelling of the effects of interdiffusion in
semiconductor nanostructures.
\pacs  	02.60.Cb, 02.70.Bf, 81.05.Ea, 81.07.St
\end{abstract}

\section{Introduction}

For the last decades, highly strained InGaAs/GaAs quantum wells
(QWs) \cite{cha09} have attracted considerable interest due to their
fundamental physical properties and their potential capabilities for the
fabrication of optoelectronic and photonic integrated circuits
(PICs) \cite{hat07,Qiu08} high- power semi-conductor diode lasers
 \cite{Aim01,Paq98} and solar cells  \cite{Xia14}. Due to its
important role in adjusting the emission properties of quantum
heterostructures, the interdiffusion process has been widely
investigated theoretically and experimentally in various systems
such as InGaAs/GaAs QW
 \cite{sck03,Mic01,Hul09,Liu11,Dix08,Has98,Mur92,Yam01} However,
the interface broadening and distortion of In-concentration profiles
result from the well-known indium surface segregation in
InGaAs/GaAs QW during the growth process. This phenomenon leads to a
significant deviation of the composition profiles from the expected
rectangular shape of QWs  \cite{Ila03}. This makes it difficult to
accurately control the emission energy through the postgrowth
intermixing process. That is why it is important to take into
consideration this effect while modelling the QW electronic structure
evolution as a function of the intermixing degree. Although In-segregation and In--Ga intermixing effects on InGaAs/GaAs
heterostructure have been widely investigated as a separate
phenomenon, less attention has been devoted to the investigation of
the In-segregation impact on the intermixing process. In the present
paper we theoretically investigate the intermixing effect for a
strained InGaAs/GaAs QW taking into consideration the changes in the
composition profile resulting from indium surface segregation during
the growth process.

\section{Theoretical considerations}

\begin{wrapfigure}{i}{0.5\textwidth}
\centerline{
\includegraphics[width=0.49\textwidth]{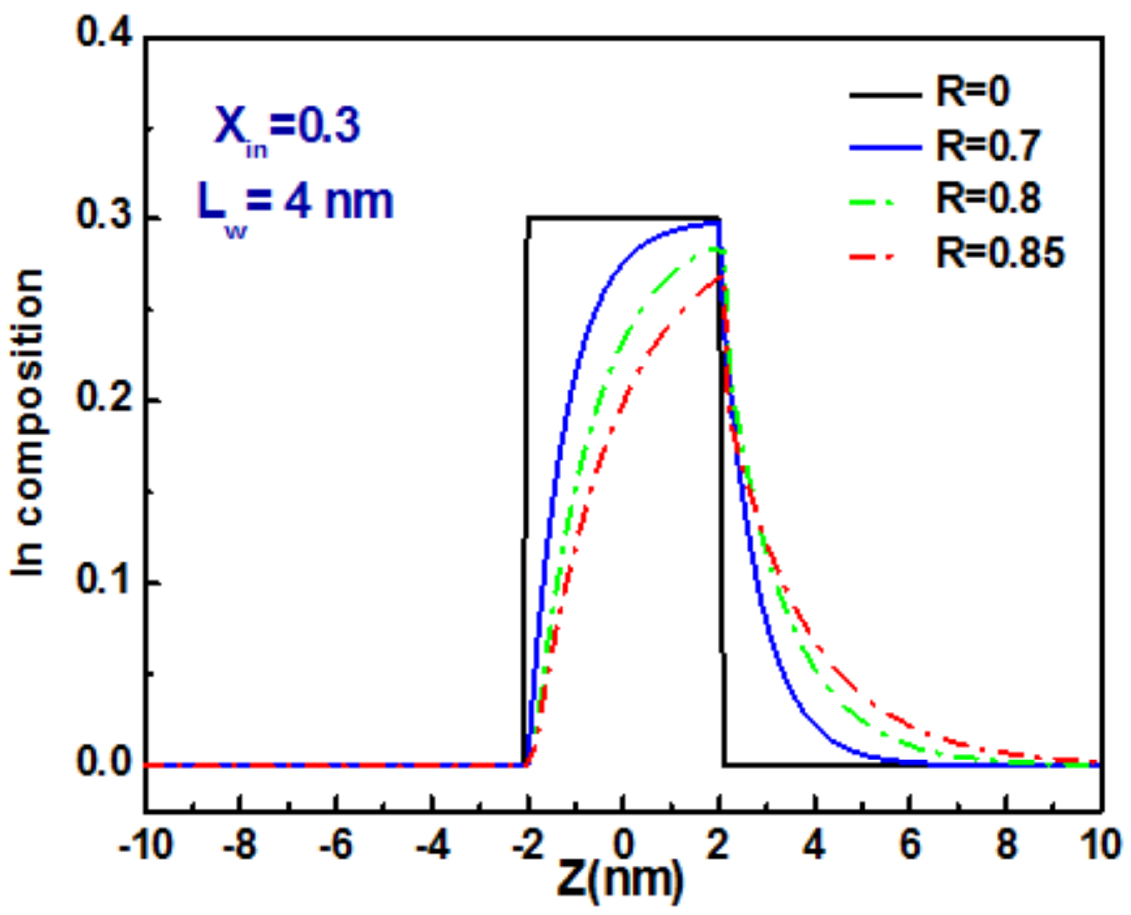}
}
\caption{(Color online) Simulated In concentration profile of an
In$_{0.3}$Ga$_{0.7}$As/GaAs QW for different segregation
coefficients using Muraki's model.} \label{F1}
\vspace{2ex}
\end{wrapfigure}

During the growth of InGaAs layer on GaAs, a strong segregation of In
atoms occurs which substantially alters the In distribution profile
along the growth direction. Therefore, the quantum well optical and
electronic properties are seriously changed by the segregation
effects. This phenomena is mainly provoked by the high In mobility
in the floating layer due to their small binding energy compared to
that in the bulk material \cite{Eva95}. This effect has been
quantified by Muraki et al.  \cite{Mur92} in a phenomenological model,
where a fraction $R$ of the deposited In atoms floats to the topmost
surface and only the remaining portion will be incorporated.
Accordingly, the In concentration in the $n$-th layer is given by:
\begin{equation}
\label{1}
 \left\{\begin{array}{lll}
x(n)= {x_0}(1-{R^n})          & \hbox{in the QW} &   0\leqslant n\leqslant N, \\
x(n)= {x_0}(1-{R^n}){R^{n-N}} & \hbox{in  the  barrier} &  n<N,
\end{array} \right.
\end{equation}
where $R$ is the segregation coefficient, $x_0$ is the nominal indium
composition and $N$ is the number of InGaAs monolayers grown before
depositing GaAs capping layer. Figure~\ref{F1} shows the impact
of the segregation coefficient on the In distribution profile of a 4
nm thick InGaAs QW with $30\%$ of In composition.

The asymmetrical In distribution profile complicates an accurate
modelling of the emission energy dependence on the intermixing
degree. Indeed, postgrowth intermixing process is highly required to
voluntarily tune the QW emission energy for several applications. In
this process, a thermally induced In/Ga atomic interdiffusion
takes place leading to the modification of the In distribution
profile and consequently alters the QW's material band gap and
the confined energy levels. It can be modelled by solving the
conventional Fick's second law of diffusion given by the equation~(\ref{2})  \cite{Chu95,Vur01}:
\begin{equation}
\label{2}
 D(T)\frac{{{\partial^2}C(z,t)}}{{\partial{z^2}}}={\frac{{\partial
C(z,t)}}{{\partial t}}}\, ,
\end{equation}
where $D(T)$ presents the diffusivity, and it can be written as a
function of the annealing temperature as follows:
\begin{equation}
\label{3}
 D(T) = {D_0}\exp\left(-\frac{E_\textrm{a}}{k_\textrm{B}{T}}\right),
\end{equation}
where $D_0$ is the diffusion coefficient, $E_\textrm{a}$ is the activation energy
for diffusion; $k_\textrm{B}$ is the Boltzmann constant, $T$ and $t$
respectively present the annealing temperature and the annealing time. The
rate of indium concentration change is proportional to the
concentration gradient $\partial C(z,t)/t$. We assume that the
diffusion coefficient is constant and all the atomic movement is
independent of atomic concentration. The interdiffusion length is
given as a function of the diffusivity and the annealing time by:
\begin{equation}
\label{4} \ L_D = \sqrt{D(T){t}}\,.
\end{equation}

\begin{wrapfigure}{i}{0.5\textwidth}
\centerline{
\includegraphics[width=0.49\textwidth]{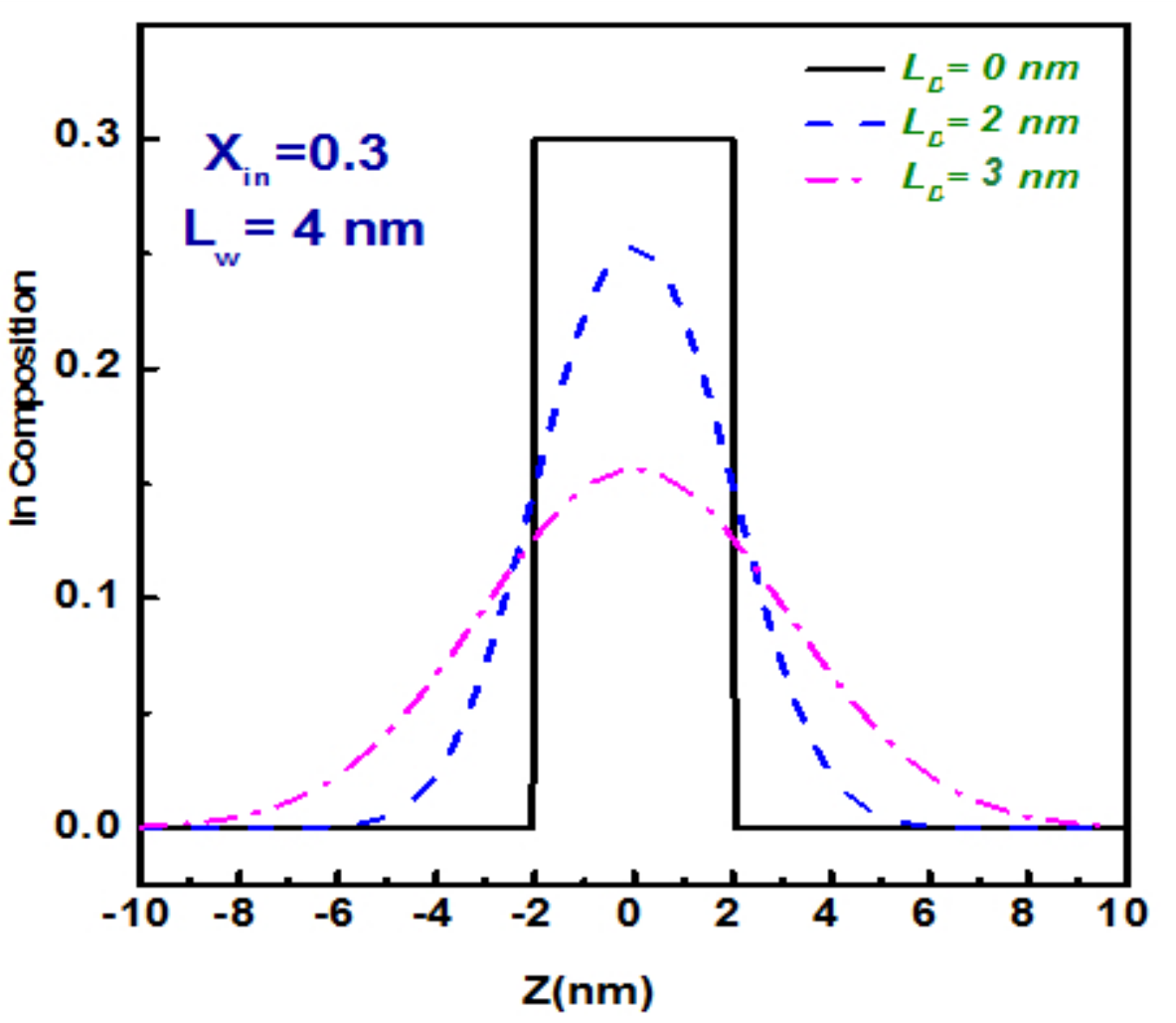}
}
\caption{(Color online)
Simulated indium concentration profiles as a function of In
diffusion length by using equation~(\ref{5}).} \label{F2}
\end{wrapfigure}

In most of the reported cases, where the segregation effect is
neglected, an approximate solution based on the error function,
given by the equation (\ref{5}), is employed to model the QW
intermixing at different diffusion lengths  \cite{Tsa96}.
\begin{equation}
\label{5}
 C(z) =\frac{{x_0}}{{2}}\left[\textrm{erf}\left(\frac{L_W-2z}{4L_D}\right)
 +\textrm{erf}\left(\frac{L_W+2z}{4L_D}\right)\right].
\end{equation}
In this relation, $C(z)$ represents  the indium concentration at
position $z$, $L_W$ is the width of the InGaAs QW, $x_0$ is the indium
mole fraction of the As-grown InGaAs layer and $L_D$ is the
diffusion length. In the absence of segregation, a conventional
rectangular QW has a symmetric In distribution profile. In this
particular case $(R=0)$, both equations (\ref{2}) and (\ref{5}) succeed
in modelling the expected In distribution profile after postgrowth
intermixing, as shown in figure~\ref{F2}, where the indium
concentration profile is given for different diffusion lengths.

To evaluate the energy level and the emission energy for a strained
In$_{0.3}$Ga$_{0.7}$As/GaAs QW and to study its dependence on the
segregation and intermixing effects, the 1D stationary
Schr\"{o}dinger equation  \cite{Ju07,Zha11} has been numerically
solved:
\begin{equation}
\label{6}
 \frac{{-{\hbar ^2}}}{2}\frac{{\partial}}{{\partial{z}}}\left[
 \frac{{1}}{m^\ast(z)}
  \frac{{\partial\psi(z)}}{{\partial{z}}}\right] + qV(z)\psi (z)= E\psi(z),
\end{equation}
where $E$ is the quantized energy level $\psi(z)$, $V(z$) and $m^\ast(z)$ are,
respectively, the wave function, the barrier potential energy, and
the position dependent carrier's effective mass in the growth
direction is chosen to be along the $z$ axis. The numerical solution has
been performed using the finite difference method within the framework of
the effective mass approximation. The unstrained band gap energy
In$_x$Ga$_{1-x}$As is given by the following formula:
\begin{equation}
\label{7}
  E_{\textrm{g},0} = x E_\textrm{g,InAs}+(1 -x)E_\textrm{g,GaAs}-x(1-x)B,
\end{equation}
where $E_\textrm{g,InAs}$ and $E_\textrm{g,GaAs}$ are the unstrained band gap of
InAs and GaAs materials and $B$ is the bowing parameter. The
considered thickness and composition of the InGaAs material is below
the critical thickness regime. Therefore, it can be considered as
coherently strained. The mismatch between In$_x$Ga$_{1-x}$As and GaAs
along the growth direction ($z$-axis) result in a compressive strain
leading to an increase in the band gap energy of the
QW \cite{Vur01}. The uniaxial strain lifts the degeneracy of the
light and heavy hole sub-bands at the centre of the brillouin zone
 \cite{Chu95}. The contribution of the hydrostatic strain to the
bulk material's band gap is given by:
\begin{equation}
\label{8}
 \varepsilon_{\bot}(x) =  2A(x)\left[1 -\frac{C_{12}(x)}{C_{11}(x)}\right]\varepsilon(x),
\end{equation}
and the change to the band gap due to the uniaxial component of the
strain is given by:
\begin{equation}
\label{9}
 \varepsilon_{\|}(x) = -b(x)\left[1 +2\frac{C_{12}(x)}{C_{11}(x)}\right]\varepsilon(x),
\end{equation}
 where $A$ and $b$ are, respectively, the hydrostatic deformation potential and shear deformation potential, $C_{12}$ and $C_{11}$ are the stiffness constants, and
 $\varepsilon$ the initial strain defined as:
\begin{equation}
\label{10}
 \varepsilon(x)=\frac{{{a_\textrm{GaAs}}-a(x)}}{{{a_\textrm{GaAs}}}}\,.
\end{equation}
  $a_\textrm{GaAs}$ and $a(x)$ are, respectively, the lattice parameter of GaAs and
  In$_x$Ga$_{1-x}$As. All the parameters of the In$_x$Ga$_{1-x}$As material are deduced from
those of  GaAs and InAs materials by using Vegard's law ensuring
that their values depend on the In compositional profile
 \cite{Vur01}. The strained band gap is given by  \cite{Tsa96}
\begin{equation}
\label{11}
  E_\textrm{g}(x) =E_{\textrm{g},0}+\varepsilon_{\bot}(x)-\varepsilon_{\|}(x).
\end{equation}
The confining potential $V(z)$ for electrons in the conduction band
and for holes in the valence band is equal to band gap discontinuity in
each band.
\begin{equation}
\label{12}
 \ V(x) = Q^i\left[E_\textrm{g,GaAs}-E_\textrm{g}(x)\right],
\end{equation}
 where $Q^i$ is the band offset taken to be $0.65$ for the conduction band and $0.35$ for the valence band.

\section{Results and discussion}

To achieve an accurate modelling of interdiffusion effect in InGaAs/GaAs QW, it is necessary to solve the exact Fick's law equation.

Accordingly,
we have applied equation (\ref{2}) for In$_{0.3}$Ga$_{0.7}$As/GaAs QW
($L_W=4$~nm) taking into account the indium segregation with an
intermixing time t of 30~s. The solution of the Fick's law allows
to investigate the effect of the intermixing process on the band
structure profile as a function of annealing temperature for
different segregation coefficients.

\begin{figure}[!h]
\centerline{
\includegraphics[width=0.7\textwidth]{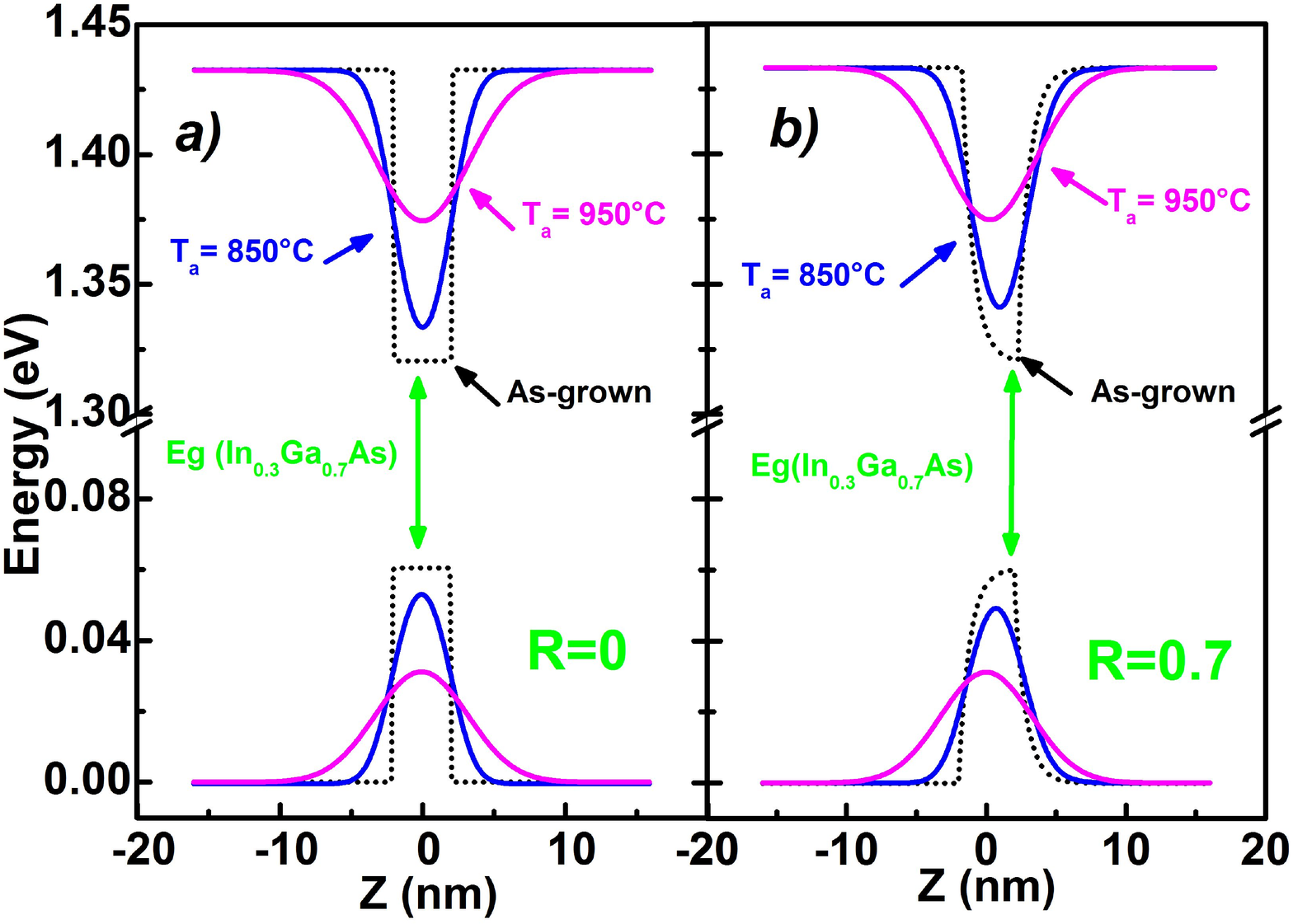}
}
\caption{(Color online) Intermixing effects on the band profile of InGaAs/GaAs QW [(a)
		without segregation and (b) with segregation] for two different
		annealing temperatures $T =850$~\textcelsius, 950~\textcelsius.} \label{F3}
\end{figure}

Figure~\ref{F3} illustrates the typical intermixing effect of
In$_{0.3}$Ga$_{0.7}$As QW at two annealing temperatures $T =850${\textcelsius}, 950{\textcelsius} for $R=0$ [figure~\ref{F3}~(a)] and $R=0.7$ [figure~\ref{F3}~(b)].
Under an annealing temperature of $T =850${\textcelsius}, the band gap
energy difference of InGaAs material arising from the segregation
effects is found to be around 30~meV. In the meanwhile, for a higher
intermixing degree, ($T= 950${\textcelsius}), the InGaAs bandgap with
segregation is only 10~meV higher than that of the reference QW
($R=0$). The energy variation induced by the segregation
effects is found to be reduced with an increase of the intermixing
degree. This behavior is likely to arise from the overall decrease
of the In concentration in the QW material  \cite{Pro01}. This
evolution also indicates that the intermixing induced energy shift
may be strongly affected by the segregation phenomenon.

\begin{figure}[!t]
\centerline{
\includegraphics[width=0.65\textwidth]{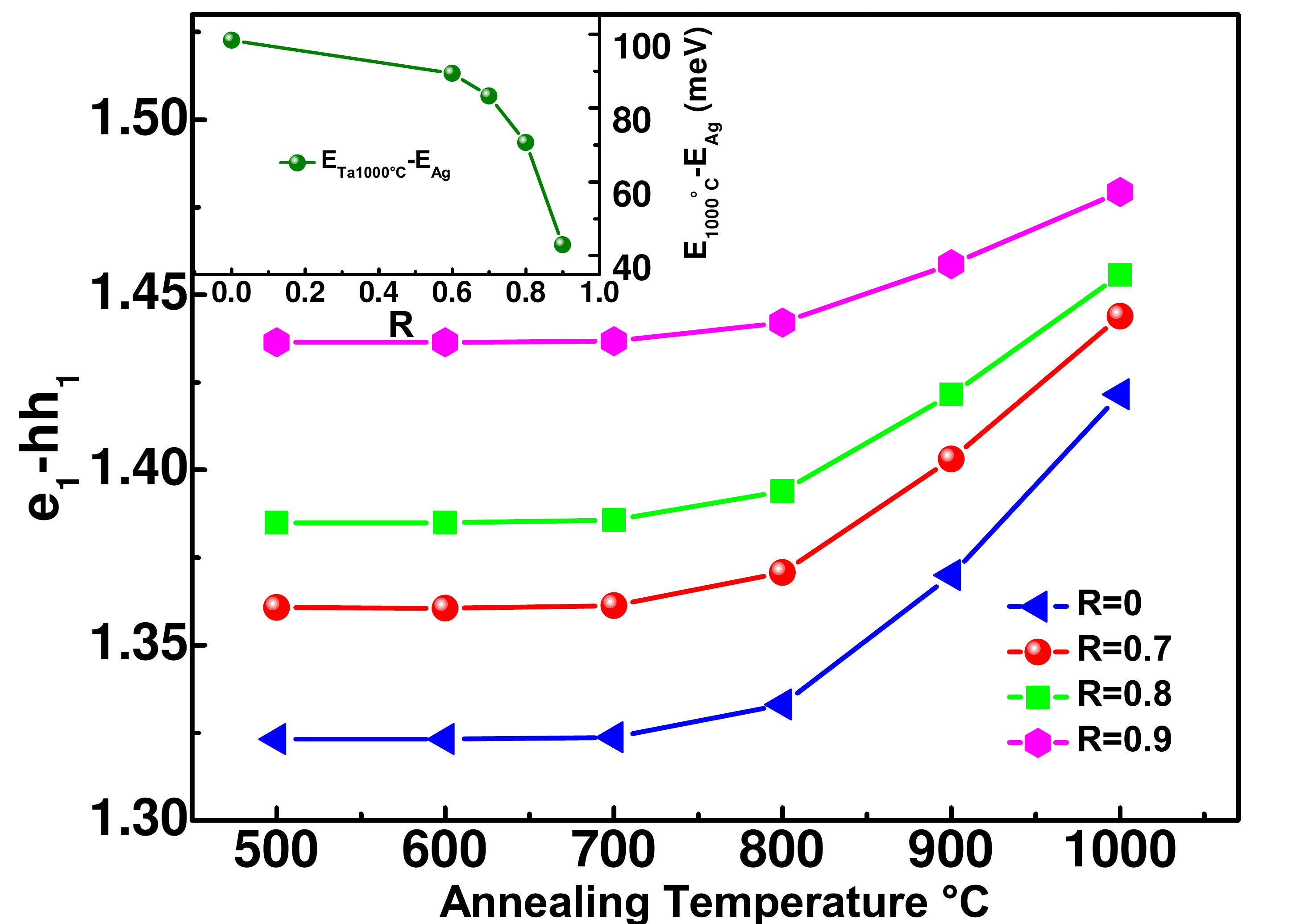}
}
\caption{(Color online) Emission energy as a function of the annealing temperature for different segregation
coefficients~$R$. Inset: Energy shift $E_{T=1000~\text{{\textcelsius}}}-E_\textrm{Ag}$ as a
function of the segregation coefficient.}\label{F4}
\end{figure}

 To illustrate this effect, we have calculated the evolution
of the emission energy shift as a function of the annealing
temperature for various values of the segregation coefficient
$R=0$, 0.7, 0.8, 0.9. The results are shown in figure~\ref{F4}, where
the energy shifts are not significantly affected for $T=800${\textcelsius}. However, for a higher annealing temperature, the results show an
important energy shift. For example at $T=1000${\textcelsius}: $E_{T=1000~\text{\textcelsius}}-E_\textrm{Ag}$ it found to be about 100~meV for $R=0$. However, it
drops down to 45~meV for a larger segregation coefficient ($R=0.9$). Indeed, the energy shift $E_{T_\textrm{a}=1000~\text{\textcelsius}}-E_\textrm{Ag}$
decreases as the segregation coefficient increases.

 This evolution can be explained
by an overall decrease in the indium composition and broadening in
the InGaAs QW. As a result, the QW becomes poorer in indium and the
interdiffusion effect becomes less sensitive to an increase of the
annealing temperature. On the other hand, for high growth
temperatures ($T_\textrm{g} \gg520${\textcelsius}), the segregation coefficient is
generally accompanied by a decrease of the point defects' density in
the vicinity of the QW  \cite{Agh11}, thus enhancing the thermal stability
of the QW structure which greatly reduces the phenomenon of
interdiffusion.

\clearpage

\section{Conclusion}

In summary, we have theoretically investigated the interdiffusion
effects in a single strained asymmetric InGaAs/GaAs QWs by numerically
solving the Fick's second law of diffusion and the one dimension
Schr\"{o}dinger equation. The calculations have been performed for
QW with segregation effects and consequent asymmetric In
distribution profile. The intermixing induced QW emission energy
shift has been found to decrease with an increase of the segregation
coefficient and interpreted in terms of overall reduction of In
concentration. This work helps to improve the understanding of the
intermixing effect in nanostructures.

\section*{Acknowledgement}

The authors would like to extend their sincere appreciation to the Deanship of Scientific Research at king Saud University for its funding this Research group No.~(RG--1436--014).

\ukrainianpart

\title{Моделювання ефектів взаємного перемішування у сильнонапруженій асиметричній
InGaAs/GaAs квантовій ямі}

\author[]%
{М. Соуаф\refaddr{label1}, М. Байра\refaddr{label1},
Г. Маареф\refaddr{label1}, Б. Ілахі\refaddr{label2,label1}}
\addresses{
\addr{label1}  Лабораторія мікроелектроніки і наноструктур, природничий факультет, Університет Монастір, Монастір,  Туніс
\addr{label2} Природничий коледж, університет короля Сауда, Ер-Ріяд, Саудівська Аравія
}

\makeukrtitle

\begin{abstract}
\tolerance=3000%
У даній роботі ми теоретично дослідили ефект взаємного перемішування  у сильнонапруженій  квантовій ямі In$_{0.3}$Ga$_{0.7}$As/GaAs, беручи до уваги зміну концентраційного профілю в результаті  {\it in-situ} поверхневої сегрегації індію.
З метою вивчення впливу ефектів сегрегації на взаємне перемішування в післяростовий період, чисельно розв'язані  одновимірне рівняння Шредінгера
в стаціонарному режимі і другий закон дифузії Фіка, використовуючи методи скінчених різниць.
Розглянуто вплив взаємної дифузії In/Ga на енергію емісії квантової ями для різних коефіцієнтів сегрегації індію. Наші результати показують, що
енергія емісії взаємноперемішаних квантових ям сильно залежить від ефектів сегрегації. Виявлено, що підсилений взаємною
дифузією енергетичний зсув
значно зменшується для вищих коефіцієнтів сегрегації.
Ця робота значно поглиблює розуміння та моделювання ефектів взаємної дифузії у напівпровідникових наноструктурах.
\keywords InGaAs/GaAs, квантові ями, моделювання, взаємна дифузія, сегрегація індію, закон Фіка

\end{abstract}


\begin{thebibliography}{99}

\bibitem{cha09} Sharma~T.K., Jangir~R., Porwal~S., Kumar~R., Ganguli~T., Zorn~M.,
Phys. Rev. B, 2009, \textbf{80}, 165403; \\ \doi{10.1103/PhysRevB.80.165403}.

\bibitem{hat07} Hatefi-Kargan~N., Steenson~D.P., Harrison~P., Linfield~E.H., Khanna~S.,
Chakraborty~S., Dean~P., Upadhya~P.C., Farrer~I., Ritchie~D.A., Sherliker~B., Halsall~M.,
Infrared Phys. Technol., 2007, \textbf{50}, 106; \\ \doi{10.1016/j.infrared.2006.10.024}.

\bibitem{Qiu08} Qiu~Y.N., Sun~H.D., Rorison~J.M., Calvez~S., Dawson~M.D., Bryce~A.C., 
    Semicond. Sci. Technol., 2008, \textbf{23}, 095010; \doi{10.1088/0268-1242/23/9/095010}.

\bibitem{Aim01} Aimez~V., Beauvais~J., Beerens~J., Ng.~S.L., Ooi~B.,
Appl. Phys. Lett., 2001, \textbf{79}, 3582; \doi{10.1063/1.1421234}.

\bibitem{Paq98} Paquette~M., Aimez~V., Bleauvais~J., Beerens~J., Poole~J.P., Charbonneau~S., Roth~A.P.,
    IEEE J. Sel. Top. Quantum Electron., 1998, \textbf{4}, No.~4, 741; \doi{10.1109/2944.720487}.

\bibitem{Xia14} Li~X.,  Dasika~V.D.,  Li~P.-Ch.,  Ji L.,  Bank~S.R.,  Yu~E.T.,
    Appl. Phys. Lett., 2014, \textbf{105}, 123906; \doi{10.1063/1.4896739}.

\bibitem{sck03} Schiettekatte~F., Aimez~V., Chicoine~M., Chevobbe~S., Chabot~J.F., Rajotte~J.F.,
    AIP Conf. Proc., 2003, \textbf{680}, 609; \doi{10.1063/1.1619790}.

\bibitem{Mic01} Chan~M.C.Y., Surya~C., Wai~P.K.A.,
J. Appl. Phys., 2001, \textbf{90}, 197; \doi{10.1063/1.1370110}.

\bibitem{Hul09} Hulko~O., Thompson~D.A., Robinson~B.J., Simmons~J.G.,
J. Appl. Phys., 2009, \textbf{105}, 073507; \doi{10.1063/1.3103332}.

\bibitem{Liu11} Liu~H.F., Liu~W., Yong~A.M., Zhang~X.H., Chua~S.J., Chi~D.Z.,
J. Appl. Phys., 2011, \textbf{110}, 063505; \doi{10.1063/1.3638703}.

\bibitem{Dix08} Dixit~V., Liu~H.F., Xiang~N.,
J. Phys. D: Appl. Phys., 2008, \textbf{41}, 115103; \doi{10.1088/0022-3727/41/11/115103}.

\bibitem{Has98}  Ghouma M., Hassen F., Sghaier~H.,  Maaref~H.,  Murray R.,
Microelectron. Eng., 1998, \textbf{43}-\textbf{44}, 197; \\ \doi{10.1016/S0167-9317(98)00164-6}.

\bibitem{Mur92} Muraki~K.,  Fukatsu~S.,  Shiraki Y., Ito R.,
Appl. Phys. Lett., 1992, \textbf{61}, 557; \doi{10.1063/1.107835}.

\bibitem{Yam01} Yamaguchi~K., Yasuda~Y., Kovacs~A., BarnavP.B.,
J. Appl. Phys., 2001, \textbf{89}, 1; \doi{10.1063/1.1331334}.

\bibitem{Ila03} Ilahi B., Sfaxi L., Bouzaiene L., Hassen F., Maaref H.,
Physica E, 2003, \textbf{17}, 232; \doi{10.1016/S1386-9477(02)00771-3}.

\bibitem{Eva95}  Evans K.R.,  Kaspi R.,  Ehret J.E.,  Skowronski M., Jones C.R.,
 J. Vac. Sci. Technol. B, 1995, \textbf{13}, 1820; \\ \doi{10.1116/1.587819}.

\bibitem{Chu95} Chuang S.L., Physics of Optoelectronic Devices, Wiley, New York, 1995.

\bibitem{Vur01} Vurgaftman I.,  Meyer J.R., Ram-Mohan  L.R.,
J. Appl. Phys., 2001, \textbf{89}, 5815; \doi{10.1063/1.1368156}.

\bibitem{Agh11} Aziz Aghchegala V.L., Mughnetsyan V.N., Kirakosyan A.A.,
Superlattices Microstruct., 2011, \textbf{49}, 99; \\ \doi{10.1016/j.spmi.2010.11.008}.

\bibitem{Tsa96} Tsang J.S.,  Lee C.P.S., Lee H.,  Tsai K.L.,  Tsai C.M.,  Fan~J.C.,
J. Appl. Phys., 1996, \textbf{79}, 664; \doi{10.1063/1.360810}.

\bibitem{Ju07}  Ju G.-X., Cai C.-Y., Xiang Y., Ren Z.-Z.,
Commun. Theor. Phys., 2007, \textbf{47}, 1001; \doi{10.1088/0253-6102/47/6/007}.

\bibitem{Zha11} Zhanga A.-P.,  Shi P., Ling Y.-W., Hua Z.-W.,
Acta Phys. Pol.~A, 2011, \textbf{120}, 987.

\bibitem{Qu04}  Qu Y., Liu C.Y., Yuan S., Wang S.Z., Yoon S.F., Chan~C.Y.,  Hong~M.H.,
J. Appl. Phys., 2004, \textbf{95}, 3422; \\ \doi{10.1063/1.1651322}.

\bibitem{Cha95} Charbonneau S., Poole P.L.,  Piva P.G.,  Aers G.C., Kotles~E.S.,
Fallahi~M., He~J.-J., Mccaffery~J.P.,  Buchanan~M., Dion~M.,  Goldberg~R.D., Mitchell~I.V., 
J. Appl. Phys., 1995, \textbf{78}, 3697; \doi{10.1063/1.359948}.

\bibitem{Pro01} Prol M., Moredo-Araujo A., Fraile-Pelaez F.J., Gomez-Alcala R., 
Superlattices Microstruct., 2001, \textbf{30}, 2; \\ \doi{10.1006/spmi.2001.0991}.

\end{thebibliography}
\end{document}